\documentclass[%
aps,nofootinbib,notitlepage,superscriptaddress,10Limberkipt,prd, twocolumn,
 amsmath,amssymb
]{revtex4-2}

\usepackage[caption=false]{subfig}
\usepackage{braket}
\usepackage{float}
\usepackage{lipsum}
\usepackage{graphicx}
\usepackage{dcolumn}
\usepackage{bm}
\usepackage{natbib}
\usepackage{hyperref}

\hypersetup{
	colorlinks = true,
    linkcolor = Maroon,
    urlcolor  = Maroon,
    citecolor = Maroon
}
\usepackage{graphicx}
\usepackage{microtype}
\usepackage{verbatim}
\usepackage[amssymb]{SIunits}
\usepackage{tabularx}
\usepackage[dvipsnames]{xcolor}
\usepackage{tikz}

\begin{document}

\preprint{APS/123-QED}

\title{Monodromic Dark Energy and DESI}

\author{Samuel Goldstein}
\email{sjg2215@columbia.edu}
\affiliation{Department of Physics, Columbia University, New York, NY 10027, USA}
\affiliation{Max-Planck-Institute für Astrophysik,  Karl–Schwarzschild–Straße 1, 85748 Garching, Germany}

\author{Marco Celoria}
\email{m.celoria@cineca.it}
\affiliation{CINECA Consorzio Interuniversitario, 40033 Casalecchio di Reno (BO), Italy}

\author{Fabian Schmidt}
\email{fabians@mpa-garching.mpg.de}
\affiliation{Max-Planck-Institute für Astrophysik,  Karl–Schwarzschild–Straße 1, 85748 Garching, Germany}

\begin{abstract}
  \noindent  Recent baryon acoustic oscillation (BAO) measurements from the Dark Energy Spectroscopic Instrument (DESI) collaboration have renewed interest in dynamical dark energy models, particularly those that cross the ``phantom divide" ($w_{\rm DE} = -1$). We present the first observational constraints on \emph{monodromic k-essence}, a physically motivated scalar field dark energy scenario capable of realizing rapid oscillations about the phantom divide.
  Using cosmic microwave background (CMB) information, DESI DR2 BAO measurements, and Type Ia supernovae observations, we constrain the amplitude, frequency, phase, and power-law index describing the monodromic k-essence scenario at the background level. We find that the monodromic dark energy scenario can fit these datasets with a $\chi^2$ that is comparable to the phenomenological $w_0$-$w_a$ parametrization. While the CMB and BAO data alone are consistent with the standard $\Lambda$CDM model, the inclusion of DESY5 supernovae shows a preference for a non-zero amplitude, $A=0.44^{+0.16}_{-0.12}$ (fully marginalized 68\% C.L.). Conversely, inclusion of the Pantheon-Plus supernovae provides no evidence for monodromic k-essence, with $A<0.43$ (95\% C.L.). We show that constraints on both monodromic dark energy and $w_0$-$w_a$ models are sensitive to the DESI DR2 LRG2 BAO distance, especially in the absence of supernovae data.
\end{abstract}

\maketitle


\section{Introduction}\label{sec:introduction}

\noindent Observations of large-scale structure (LSS), Type Ia supernovae, and the cosmic microwave background (CMB) have firmly established that the Universe is undergoing a period of accelerated expansion~\cite{Efstathiou:1990xe, White:1993wm, SupernovaSearchTeam:1998fmf, SupernovaCosmologyProject:1998vns, WMAP:2003ivt, Percival:2007yw}. Uncovering the physical mechanism responsible for this accelerated expansion, termed dark energy, remains one of the most significant open questions in cosmology. The standard cosmological model, $\Lambda$CDM, attributes this accelerated expansion to the cosmological constant ($\Lambda$). However, the cosmological constant faces several theoretical challenges, most notably the large discrepancy between the observed value of the vacuum energy and theoretical expectations~\cite{Weinberg:1988cp}, leading to the supposition that the cosmological constant in fact vanishes, and the acceleration is driven by a different source instead. In particular, a wide range of dynamical dark energy models have been proposed~\cite{Frieman:2008sn,Li:2011sd}, for which the dark energy density typically evolves over cosmic time.

Recently, baryon acoustic oscillation (BAO) measurements from the Dark Energy Spectroscopic Instrument (DESI) collaboration have revealed mild tensions within the standard $\Lambda$CDM model, particularly when compared with Type Ia supernovae measurements~\cite{DESI:2024mwx,DESI:2025zgx}. These tensions have sparked significant interest in dynamical dark energy. Interestingly, non-parametric reconstructions of the dark energy equation of state using BAO and Type Ia supernovae observations have shown hints of oscillations about the ``phantom divide," $w_{\rm DE}=-1$~\cite{Zhao:2017cud,Berti:2025phi, DESI:2025fii, DESI:2025wyn}. Although caution is warranted in over-interpreting data-driven reconstructions that favor oscillations precisely about $w_{\rm DE}=-1$, and although such phantom crossings can pose theoretical challenges at the level of perturbations~\cite{Kunz:2006wc} (see also the discussion below), it is interesting to explore oscillatory dark energy scenarios.

\begin{figure*}[!t]
\centering
\includegraphics[width=0.99\linewidth]{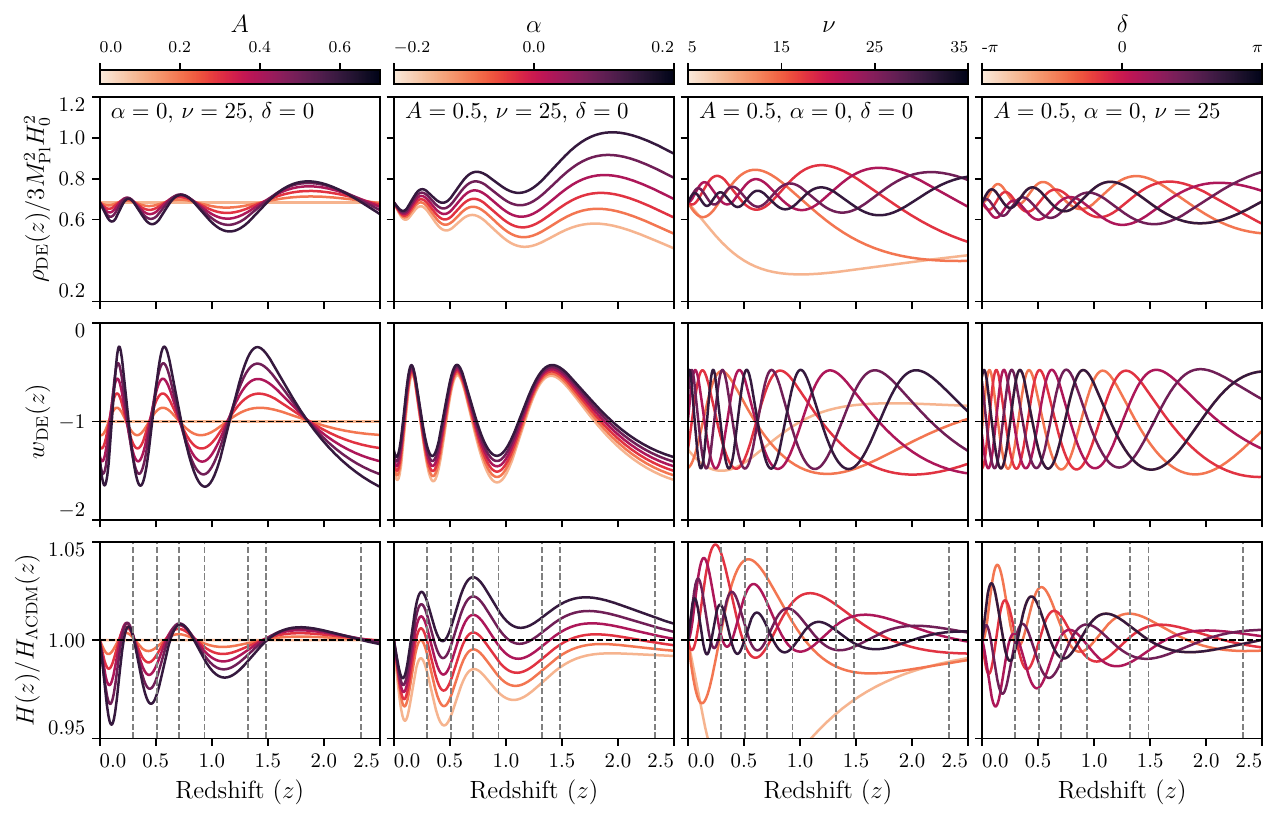}
\caption{
Impact of monodromic dark energy parameters on the evolution of the dark energy density (top), dark energy equation of state (middle), and Hubble parameter (bottom), as a function of redshift. Each column illustrates variations in a single parameter:  amplitude ($A$), power-law index ($\alpha$), frequency ($\nu$), and phase ($\delta$), from left to right. All parameters are varied across the prior ranges used in our fiducial analysis, except for the frequency, which varies across a broader range than our fiducial prior ($15\leq \nu \leq 30$) to better illustrate the impact of low and high frequency oscillations. The dashed vertical lines in the bottom panels indicate the redshifts of the DESI DR2 BAO data. The monodromic dark energy scenario is able to capture a range of oscillatory features in the cosmological expansion history.}

\label{fig:MDE_theory_predictions}
\end{figure*}

Previous works have placed observational constraints on a range of oscillatory dark energy scenarios~\cite{xia/etal:2005, barenboim1,barenboim2,linder:2006,kurek1,kurek2,pace/etal:2011, Escamilla:2024fzq, Kessler:2025kju, Gialamas:2025pwv}. These studies typically adopt phenomenological parameterizations of the dark energy equation of state, and focus on oscillations on the order of the Hubble time. However, faster oscillations are theoretically allowed and observationally viable, corresponding to a new window in the dark energy model space that has been largely unexplored \cite{Schmidt:2017iap}. Here, we place constraints on \emph{monodromic k-essence} --- a physically motivated dark energy scenario proposed by~\citet{Schmidt:2017iap} that is capable of realizing rapid oscillations in the dark energy equation of state.

To understand how monodromic dark energy can achieve fast oscillations in the dark energy evolution in a physically motivated framework, note that an approximately flat potential, which is needed for an equation of state $w_{\rm DE}$ close to $-1$, can be realized in a technically natural way by introducing a shift symmetry $\phi\to \phi+c$ which is weakly broken. In the well-known explicit constructions of inflation in the context of string theory \cite{Silverstein:2008sg}, known as \emph{axion monodromy}, nonperturbative effects lead to periodic modulations of the potential. Whereas oscillatory quintessence potentials have been considered in \cite{dodelson/kaplinghat/stewart:2000,frieman/etal:1995,kamionkowski/pradler/walker,damico/etal:2016}, a canonical scalar field is not the only possible option. Models with a non-standard kinetic term, referred to as \emph{k-essence}, are also viable \cite{chiba/okabe/yamaguchi,ArmendarizPicon:2000ah}.\footnote{Non-standard kinetic terms also appear in the context of string theory in form of the Dirac-Born-Infeld action \cite{silverstein/tong,alishahiha/etal:2004}.} Notably, \citet{Schmidt:2017iap} pointed out that k-essence scenarios allow for much more interesting oscillating phenomenology than a canonical scalar field, which, when slowly rolling, will asymptote to $w_{\rm DE}=-1$ as soon as it encounters a local minimum in the potential. This rich phenomenology motivates the present work, which confronts monodromic k-essence with cosmological data.

Given the challenges of perturbative stability of phantom dark energy models (see Sec.~\ref{sec:monodromic_dark_energy} and Appendix~\ref{app:kess_stab} for a discussion and possible solution), we constrain monodromic k-essence solely at the background level. Specifically, we constrain monodromic k-essence using a compressed CMB likelihood marginalized over late-time physics, combined with BAO and Type Ia supernovae observations. We place the first constraints on the amplitude, frequency, phase, and power-law index of the monodromic dark energy potential (Fig.~\ref{fig:MDE_theory_predictions} illustrates the phenomenological effects of these parameters). We show that the physically motivated monodromic dark energy model can fit these datasets with comparable $\chi^2$ to the widely-used, phenomenological Chevallier-Polarski-Linder (CPL) $w_0$-$w_a$ parametrization~\cite{Chevallier:2000qy,Linder:2002et}, albeit with more free parameters. We quantify the ``preference" for monodromic dark energy for several dataset combinations, and underscore the role of the DESI LRG2 data in driving dynamical dark energy constraints, for both monodromic dark energy and $w_0$-$w_a.$

The remainder of this paper is organized as follows. In Sec.~\ref{sec:monodromic_dark_energy}, we provide the theoretical background for monodromic dark energy. In Sec.~\ref{sec:datasets_and_methodology}, we describe the datasets and methodology used in this work. We present our results in Sec.~\ref{sec:results} and conclude in Sec.~\ref{sec:conclusion}. In Appendix~\ref{app:kess_stab}, we discuss an improved monodromic k-essence construction designed to mitigate instabilities in the dark energy perturbations. In Appendix~\ref{app:DESI_LRG2}, we present a detailed investigation of the impact of the DESI LRG2 BAO measurements on constraints on monodromic dark energy and $w_0$-$w_a$ models. In Appendix~\ref{app:relaxed_frequency_prior}, we present constraints on monodromic k-essence with a broader frequency prior.

\section{Monodromic dark energy}\label{sec:monodromic_dark_energy}

\noindent In this section, we review the main aspects of the monodromic dark energy model. We refer the reader to Ref.~\cite{Schmidt:2017iap} for more details. Our starting point is a spatially flat universe with a scalar field $\phi$ that drives the late-time accelerated expansion. To accommodate a rich set of expansion histories, including phantom crossing, we consider a field which drives accelerated expansion via its \emph{kinetic} energy, often referred to as k-essence~\cite{Armendariz-Picon:1999hyi,Chiba:1999ka}.\footnote{This is in contrast to the more familiar quintessence scenario, where the accelerated expansion is driven by the potential of a canonical scalar field. It is straightforward to construct a monodromic quintessence model~\cite{Schmidt:2017iap}; however, this model cannot cross the phantom divide, and is therefore of less observational interest given the dark energy scenarios preferred by DESI.} To this end, we consider the following action,
\begin{equation}
  S=\int d^4x\sqrt{-g}\left[\frac{1}{2}M_{\rm Pl}^2R+p(\phi,X)+\mathcal{L}_m \right],
  \label{eq:MDE_action}
\end{equation}
where $M_{\rm Pl}^2 \equiv 1/(8\pi G_{\rm N})$ is the reduced Planck mass, $p(\phi,X)$ is a function of the scalar field, $\phi$, and its kinetic energy, $X\equiv -\frac{1}{2}g^{\mu\nu}\partial_\mu \phi\partial_\nu\phi$. For convenience, we normalize the field such that $\phi$ has units of time, or inverse mass, and hence $X$ is dimensionless. We focus on a pure k-essence model up to quadratic order in the kinetic energy, \emph{i.e.},\footnote{Any $p(\phi, X)=K(\phi)X+L(\phi)X^2$ with $L(\phi)\neq 0$ can be cast into this form via a field redefinition~\cite{Chiba:1999ka}.}
\begin{equation}
    p(\phi,X)=V(\phi)[-X+X^2],
\end{equation}
where the ``potential" is given by
\begin{equation}\label{eq:MDE_potential}
    V(\phi)=C\left(\frac{\phi}{\phi_0}\right)^{-\alpha}\left[1-A\sin(\nu {H_0} \phi+\delta)\right].
\end{equation}
Here, $A$, $\nu$, and $\delta$ set the amplitude, frequency, and phase of the oscillations, respectively. We have introduced the Hubble rate today $H_0$ in order to make $\nu$ dimensionless; very roughly, $\nu$ sets the number of oscillations executed over the age of the Universe.
The power-law index $\alpha$ specifies the smooth component of the potential, and is therefore related to the time-averaged equation of state of the field. In the most general construction, $\phi_0$ and $C$ are free parameters; however, as shown in Ref.~\cite{Schmidt:2017iap}, the monodromic k-essence scenario admits a tracking solution during matter domination when $A=0$. In this work, we initialize the field on this tracking solution, in which case $C$ and $\phi_0$ are fully determined by the present-day dark energy density. See~Ref.~\cite{Schmidt:2017iap} for the precise form of this tracking solution.

To better understand how the parameters in Eq.~\eqref{eq:MDE_potential} influence the cosmological expansion history, Fig.~\ref{fig:MDE_theory_predictions} shows how variations in each parameter individually affect the dark energy density (top), equation of state (middle), and Hubble parameter (bottom). From left to right, we vary the amplitude $A$, power-law index $\alpha$, frequency $\nu$, and phase $\delta$, holding all other parameters fixed at their fiducial values of $A=0.5$, $\alpha=0$, $\nu=25$, and $\delta=0$. We fix $H_0$, $\Omega_m$, and $\Omega_b$, thus anchoring the present-day dark energy density and expansion rate. All parameters except the frequency are varied across the prior ranges used in our fiducial analysis. For the frequency, we show variations across a broader range of values than our fiducial prior ($15\leq \nu \leq 30$) to highlight values of $\nu$ that are either too small to produce significant oscillations since dark energy domination, or too high to be resolved given the redshift binning of the DESI DR2 dataset (indicated by the vertical gray dashed lines).

Each parameter introduces distinct features in the background history: increasing $A$ enhances the amplitude of oscillations in $\rho_{\rm DE}(z)$ and $w_{\rm DE}(z)$, while variations in $\alpha$ shift the time-averaged equation of state and tilt the time-evolution of the dark energy density. Changes in $\nu$ and $\delta$ modulate the frequency and phase of $w_{\rm DE}(z)$, respectively. By varying all of these parameters, the monodromic k-essence model can accommodate a broad range of oscillatory features in the dark energy evolution. 

Thus far, our discussion has focused solely on the impact of monodromic k-essence at the background level. However, dark energy models that cross the phantom divide generally violate the null-energy condition (NEC). This is problematic, since such violations are often accompanied by gradient instabilities, which happen if the squared sound speed of dark energy perturbations becomes negative, leading to exponential growth of perturbations on all scales~\cite{Hsu:2004vr, Vikman:2004dc}. Ref.~\cite{Creminelli:2008wc} argued that these instabilities can be kept under control by adding higher-derivative terms to the action, which are generically expected in a theory with a finite cutoff. Gradient instabilities arise in the monodromic k-essence model for a vast range of the parameter space considered here. Although including higher-derivative terms could, in principle, stabilize the theory, a complete treatment of perturbations in mondromic dark energy is left to a future work. Here, we conservatively constrain the monodromic k-essence model using observations that are insensitive to dark energy perturbations. In Appendix~\ref{app:kess_stab}, we present a modification to the monodromic k-essence model that can mitigate gradient instabilities while allowing for oscillations of similar amplitude in the equation of state of dark energy.


%
\section{Datasets and Methodology}\label{sec:datasets_and_methodology}

\noindent In this section, we describe the datasets and analysis choices used to constrain the monodromic k-essence model. Given the theoretical uncertainties and model-dependence of modeling dark energy perturbations for models that cross the phantom divide, we constrain the monodromic k-essence scenario including only information at the background level. In particular, we do not use a full CMB likelihood, which encodes information about dark energy perturbations through the integrated Sachs–Wolfe effect~\cite{Sachs:1967er} and gravitational lensing. Instead, following the DESI collaboration~\cite{DESI:2025zgx}, we adopt a compressed CMB likelihood constructed from the late-time marginalized CMB analysis of Ref.~\cite{Lemos:2023xhs}, which utilizes the Planck PR4 \texttt{CamSpec} likelihood \cite{Planck:2020olo, Rosenberg:2022sdy}. This compressed likelihood, which we denote by $Q_{\rm CMB}$, is modeled as a multivariate Gaussian in the acoustic angular scale $\theta_*$, the physical baryon density $\omega_b$, and the combined physical baryonic and cold dark matter density $\omega_{bc} = \omega_b + \omega_c$. For further details, we refer the reader to Appendix A of Ref.~\cite{DESI:2025zgx}.

\begin{figure*}[!t]
\centering
{\includegraphics[width=0.99\linewidth]{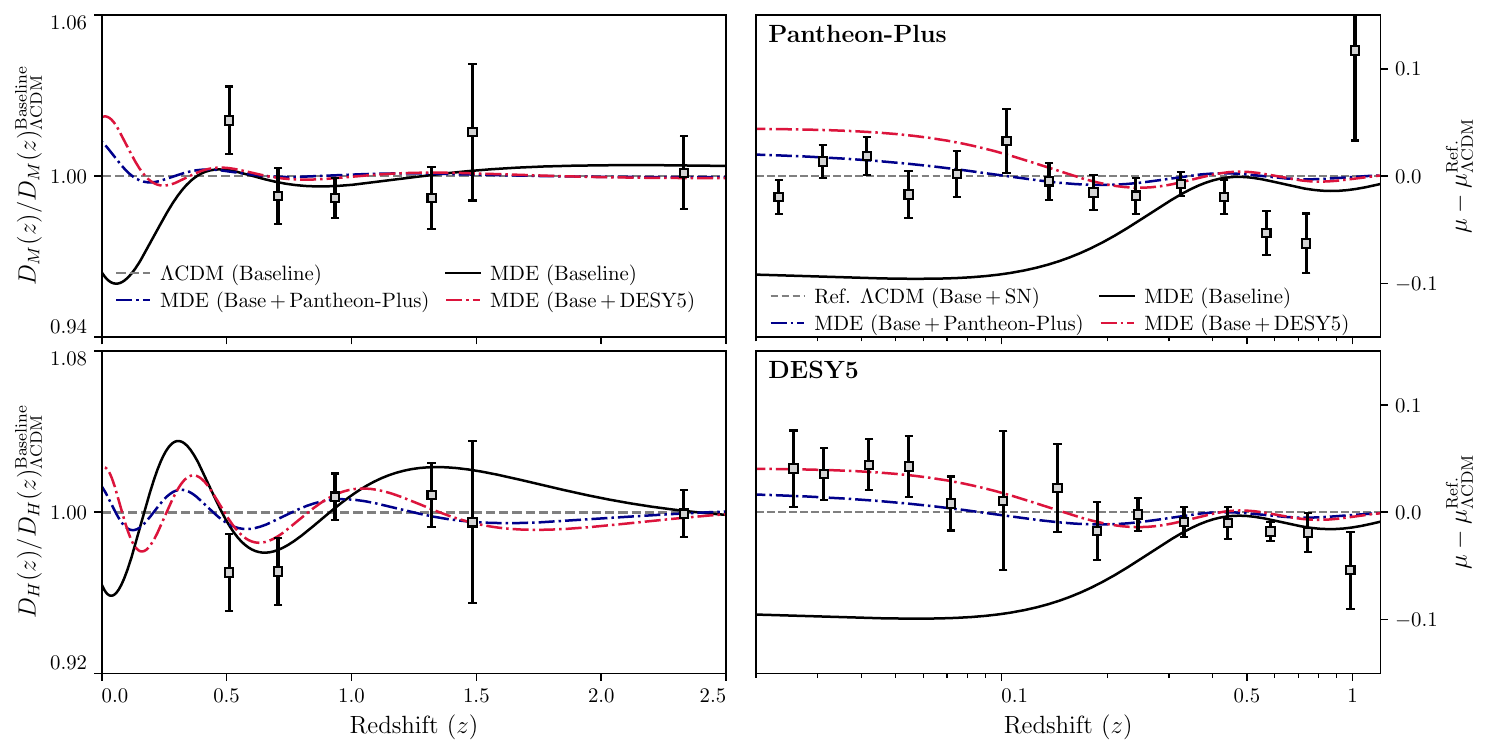}}
\caption{
  \textit{Left panel:} Comparison of the transverse (top) and parallel (bottom) BAO distance measurements for $\Lambda$CDM and monodromic dark energy models. Squares indicate DESI DR2 measurements. Measurements and theoretical predictions are plotted relative to the baseline ($Q_{\rm CMB}$+DESI DR2 BAO) best-fit $\Lambda$CDM model. The black, blue, and red lines show best-fit predictions for monodromic dark energy models using the baseline dataset alone, and in combination with Pantheon-Plus and DESY5 supernovae data, respectively.
\textit{Right panel:} Comparison of supernovae distance moduli for $\Lambda$CDM and monodromic dark energy models. For clarity, we show distance moduli relative to the best-fit $\Lambda$CDM cosmology from the baseline ($Q_{\rm CMB}$+DESI DR2 BAO) dataset combined with the particular supernovae sample shown in each panel: Pantheon-Plus (top) and DESY5 (bottom). Gray squares indicate supernovae measurements, which have been combined using inverse-variance weighting for visualization. As detailed in the text, the absolute magnitude calibration is fixed to the best-fit value obtained by fitting the unbinned supernovae data under the reference $\Lambda$CDM model in each panel.}
\label{fig:BAO_SN_measurement_comparison}
\end{figure*}

In addition to the $Q_{\rm CMB}$ data, our ``baseline" dataset includes BAO distance measurements from the second data release of the DESI survey~\cite{DESI:2025zpo,DESI:2025zgx}. Specifically, we use the combined BAO sample, which spans redshifts $0.295\leq z\leq 2.33$ and includes measurements from galaxies, quasars, and the Lyman-$\alpha$ forest. In our fiducial analysis, we use the entire DESI DR2 BAO dataset. In Appendix~\ref{app:DESI_LRG2}, we explore the impact of removing the parallel and transverse BAO distance measurements for the DESI LRG2 data point with an effective redshift $z_{\rm eff}=0.706,$ which partially drive the monodromic dark energy constraints, particularly for the baseline dataset (see the left panel of Fig.~\ref{fig:BAO_SN_measurement_comparison}).

In addition to our baseline dataset, we include uncalibrated Type Ia supernovae distance moduli measurements, which constrain the relative expansion history. We consider two supernovae datasets. First, we use the Pantheon-Plus dataset, which consists of 1550 spectroscopically-confirmed Type Ia supernovae between redshifts $0.01\leq z\leq 2.26$~\cite{Brout:2022vxf}. Second, we use the Dark Energy Survey Y5 (DESY5) supernovae dataset~\cite{DES:2024jxu}, which includes 1635 photometrically classified Type Ia supernovae spanning $0.1\leq z\leq1.3$, along with 194 low-redshift Type Ia supernovae between $0.025\leq z\leq 0.1$~\cite{DES:2024jxu}. We do not analyze the Union3 supernovae compilation~\cite{Rubin:2023ovl}, which tends to favor dynamical dark energy models lying between those preferred by the Pantheon-Plus and DESY5 samples~\cite{DESI:2025zgx, DESI:2025wyn}.

We compute theoretical predictions in the monodromic dark energy scenario using a modified version of the \texttt{hi\_class} Einstein-Boltzmann code\footnote{\href{http://miguelzuma.github.io/hi_class_public/}{http://miguelzuma.github.io/hi\_class\_public/}}~\cite{Zumalacarregui:2016pph}. \texttt{hi\_class} is an extension of the CLASS code~\cite{Lesgourgues:2011re} that incorporates Horndeski's theory~\cite{Horndeski:1974wa}, of which k-essence is a special case, at both the background and linear perturbation levels; again, we only use the background information in this work. Since varying the sum of the neutrino masses does not significantly impact $w_0$-$w_a$ constraints for the datasets considered here~\cite{DESI:2025zgx}, we assume a minimal neutrino mass $\sum m_\nu=0.06$ eV with two massless and one massive eigenstate.

We sample from the parameter posterior using \texttt{Cobaya}~\cite{Torrado:2020dgo}, assuming the following uniform priors on the monodromic dark energy parameters: $0.0\leq A\leq 0.7$, $-0.2\leq \alpha \leq 0.2$, $15\leq \nu \leq 30$, and $-\pi\leq \delta < \pi.$ Our \emph{informative} prior on frequency is motivated by two considerations. We want the frequency to be high enough to ensure that there are several oscillations in the Hubble parameter at redshifts $z\lesssim 2$, but not so high that the oscillations become unresolvable given the redshift binning of the DESI data (see Fig.~\ref{fig:MDE_theory_predictions}). We present constraints for a broader range of frequencies ($5\leq \nu \leq 35$) in Appendix~\ref{app:relaxed_frequency_prior}. We set the initial conditions, including $C$ and $\phi_0$, using the tracking solution discussed in Sec.~\ref{sec:monodromic_dark_energy}.\footnote{There is a slight subtlety in setting the initial conditions. We use the tracking solution defined during matter domination; 
however, in \texttt{hi\_class}, we have to initialize the field much earlier ($a\approx 10^{-14}$). We have verified that this initialization procedure leads the field to the attractor solution well before dark energy domination. Indeed, we find that this procedure yields late-time ($z<3$) predictions that are entirely consistent with direct numerical integration of the background equations with initial conditions consistently fixed to the tracking solution during matter domination ($a\approx 5 \times 10^{-4})$.  } We also assume the following wide uniform priors on cosmological parameters: $50\leq H_0\leq 85$ km/s/Mpc, $0.09\leq \omega_{\rm cdm}\leq 0.15$, and $0.017\leq \omega_{\rm b}\leq 0.027$. We assess convergence using the Gelman-Rubin statistic~\cite{gelman1992} with a tolerance of $R-1<0.01.$ We determine the \emph{maximum a posteriori} (MAP) parameters using the \textsc{BOBYQA} minimizer implemented in Cobaya~\cite{Cartis:2018jxl, Cartis:2018xum}.

%

\begin{figure*}[!t]
\centering
\includegraphics[width=0.99\linewidth]{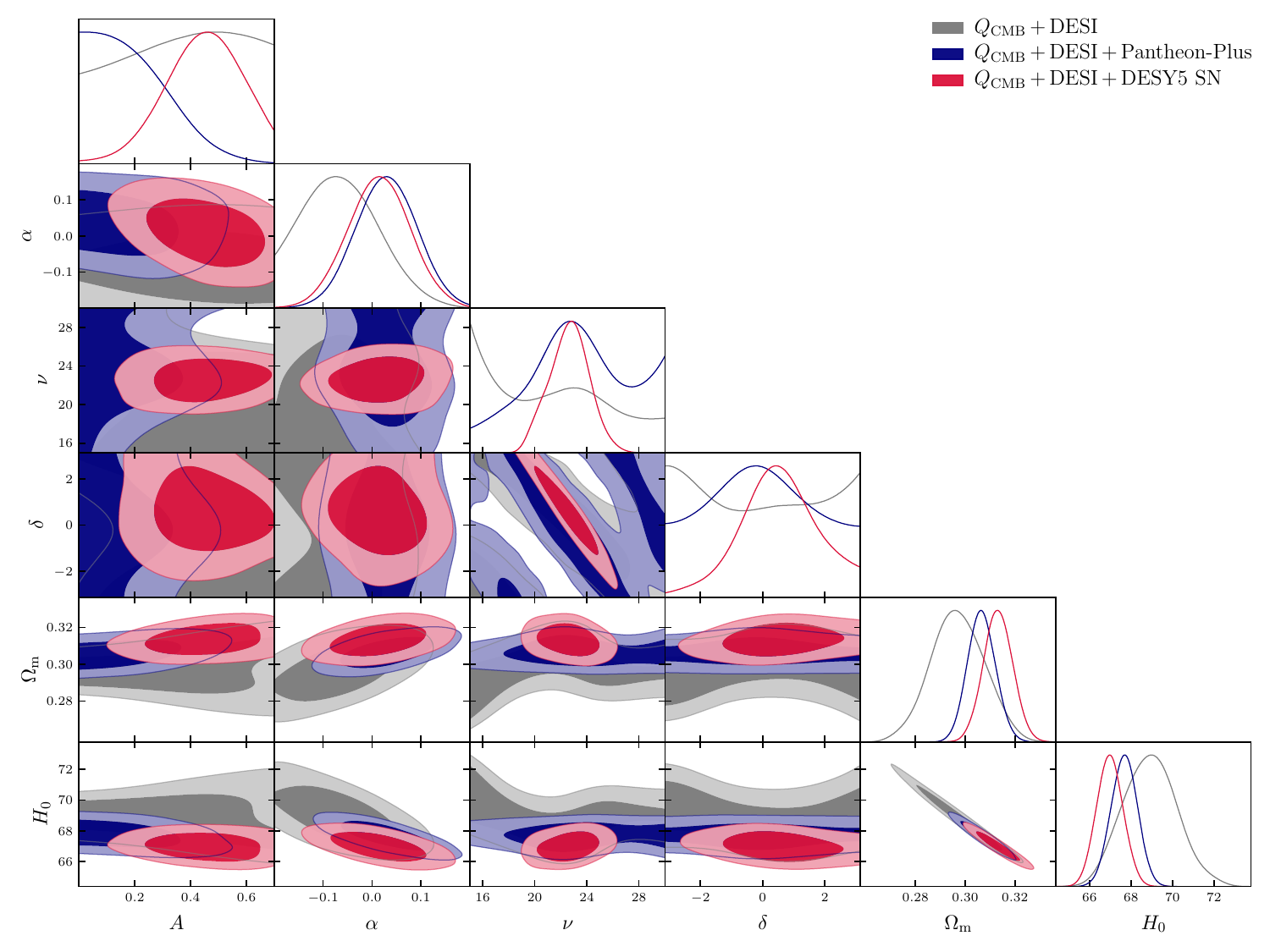}
\caption{Marginalized posterior distributions for the key monodromic dark energy and other cosmological parameters analyzed in this work. Constraints are shown for datasets combining \emph{Planck} CMB and DESI BAO (gray), with the addition of Pantheon-Plus supernovae (blue), and DESY5 supernovae (red). }
\label{fig:triangle_plot}
\end{figure*}
%


%
\begin{figure*}[!t]
\centering
\includegraphics[width=0.99\linewidth]{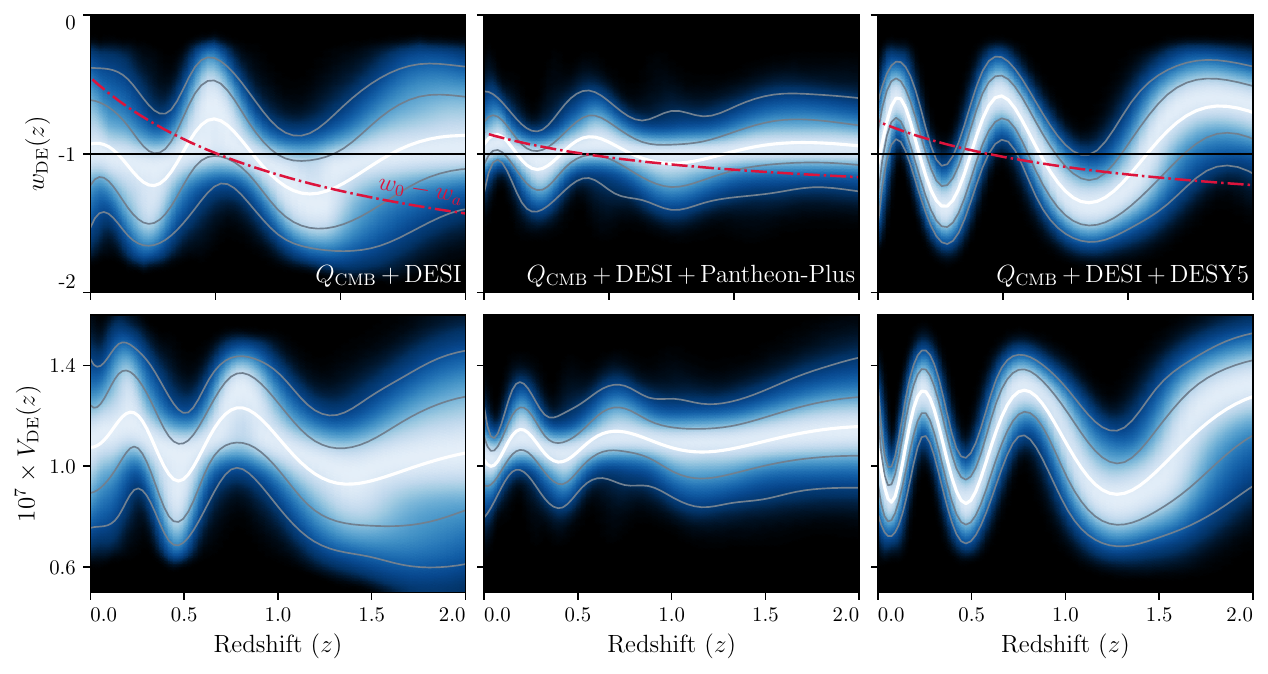}
\caption{Constraints on the redshift evolution of the monodromic k-essence equation of state (top) and potential (bottom) for the datasets considered in this work. White lines show the posterior mean, while gray bands indicate the 68\% and 95\% confidence levels. For reference, the solid black line indicates $w_{\rm DE} = -1$, and the red dot-dashed line denotes the best-fit $w_0–w_a$ model for each dataset combination. Note that the oscillatory features in this plot are partially driven by our informative prior on the frequency.}
\label{fig:reconstruction}
\end{figure*}
%


%
\section{Results}\label{sec:results}

\noindent Before analyzing the full constraints on the monodromic dark energy scenario, we compare predictions from monodromic dark energy with BAO and supernovae observations. Fig.~\ref{fig:BAO_SN_measurement_comparison} (left panel) compares predictions of the transverse (top) and parallel (bottom) BAO distance estimates for $\Lambda$CDM and monodromic dark energy models with DESI DR2 measurements. We show measurements and predictions relative to the best-fit $\Lambda$CDM prediction to the baseline dataset, consisting of $Q_{\rm CMB}$ and DESI BAO. The solid curves show the best-fit monodromic k-essence predictions for the three dataset combinations considered here. The best-fit parameters are listed in Table~\ref{tab:posterior_param_limits}. Notice the stronger oscillations in $D_H$ compared to $D_M$: this is because the former involves one less time integral over the equation of state \cite{Schmidt:2017iap}.

From the lower left panel of Fig.~\ref{fig:BAO_SN_measurement_comparison}, it is clear that the best-fit k-essence model to the baseline dataset is largely driven by the LRG2 parallel BAO data point at redshift $z\approx0.7.$ However, this particular model is disfavored once we include Type Ia supernovae measurements, which severely constrain the low redshift expansion history. Indeed, this can be seen in the right panel (solid black lines), which shows the supernovae distance moduli for the Pantheon-Plus (top) and DESY5 (bottom) datasets. For clarity, we show the inverse-variance-weighted binned supernovae distance moduli relative to the best-fit $\Lambda$CDM prediction for the particular baseline+SN dataset shown in each panel. That is, the top (bottom) panel shows results relative to the best-fit baseline+Pantheon-Plus (DESY5) $\Lambda$CDM cosmology. As the supernovae data are uncalibrated, we need to adopt an absolute magnitude calibration for visualization purposes. To this end, we compute the best-fit (MAP) absolute magnitude for each supernovae dataset by fitting the unbinned supernovae distance moduli at fixed cosmology, assuming the best-fit reference $\Lambda$CDM cosmology used in each panel.

\renewcommand{\arraystretch}{1.4} 
\begin{table*}[!t]
    \centering
    \scriptsize
     \setlength{\tabcolsep}{2pt} 

     \begin{tabular}{ |c|c|c|c|c|c|c|c| } 
        \hline
        & $A$ & $\alpha$ & $\nu$ & $\delta$ & $\Omega_m$ & $H_0$ [km/s/Mpc] & $\Delta \chi^2$  \\ 
        \hline
        Base: $Q_{\rm CMB}$+DESI 
        & ---$~(0.58)$ 
        & $-0.06^{+0.06}_{-0.09}~(-0.08)$ 
        & --- $~(15.2)$ 
        & --- $~(-2.2)$ 
        & $0.297\pm 0.011~(0.281)$ 
        & $68.9\pm 1.3~(71.0)$  
        & -7 \\ 
        \hline
        Base+Pantheon-Plus 
        & $<0.44~(0.25)$ 
        & $0.03\pm 0.06~(0.01)$ 
        & --- $~(22.1)$ 
        & --- $~(0.2)$ 
        & $0.307\pm 0.006~(0.308)$ 
        & $67.7\pm 0.6~(67.5)$  
        & -5 \\ 
        \hline
        Base+DESY5 SN 
        & $0.44^{+0.15}_{-0.12}~(0.47)$ 
        & $0.01\pm 0.06~(0.00)$ 
        & $22.6^{+1.6}_{-1.5}~(22.8)$ 
        & $0.5\pm 1.2~(0.5)$ 
        & $0.313\pm 0.006~(0.314)$ 
        & $67.0\pm 0.6~(66.9)$  
        & -16  \\ 
        \hline
    \end{tabular}

    \caption{Marginalized constraints on monodromic dark energy and other cosmological parameters for the datasets considered in this work. For each dataset, we report the posterior mean and 68\% two-tailed C.L. for all parameters that are detected at $>2\sigma$, otherwise we report the one-tailed 95\% C.L.; ``---" denotes parameters that are entirely constrained by the prior at $2\sigma.$ Maximum \textit{a posteriori} values are shown in parentheses. The final column reports the $\Delta \chi^2$ values computed with respect to a $\Lambda$CDM cosmology.
    }

   \label{tab:posterior_param_limits}
\end{table*}

Fig.~\ref{fig:triangle_plot} shows the marginalized posteriors for the key monodromic dark energy and $\Lambda$CDM parameters analyzed in this work. For the baseline dataset (gray), the constraints on the amplitude, frequency, and phase of oscillations are largely prior dominated. However, the degeneracy between the frequency ($\nu$) and the phase ($\delta$) indicates that this dataset combination has a preferred oscillation scale. As discussed in detail in Appendix~\ref{app:DESI_LRG2}, this preference is primarily driven by the ``dip" in the LRG2 BAO distance measurements compared to $\Lambda$CDM.

Our constraints that include supernovae data are sensitive to the particular supernovae sample. While the baseline+Pantheon-Plus dataset (blue) is consistent with the $\Lambda$CDM limit ($A=0$), the baseline+DESY5 dataset (red) suggests a mild preference for a non-zero amplitude, $A=0.44^{+0.15}_{-0.12}$ at the 68\% C.L. These findings are consistent with previous studies of dynamical dark energy with supernovae data, which have found that DESY5 tends to exhibit a stronger preference for evolving dark energy than Pantheon-Plus.

We also find that including supernovae data leads to a preference for a specific oscillation frequency, $\nu\approx 23$, although we emphasize that this is partially driven by our informative prior on the frequency, $15\leq \nu \leq 30$. As shown in Appendix~\ref{app:relaxed_frequency_prior}, relaxing this prior to $5 \leq \nu \leq 35$ yields a multimodal posterior, indicating that current data are compatible with a broad range of oscillation frequencies. Finally, we note that the combination of the CMB, BAO, and supernovae place strong constraints on $\alpha$, favoring oscillations about $w_{\rm DE}=-1$ ($\alpha=0)$.

To quantitatively assess if there is a preference for monodromic k-essence over $\Lambda$CDM, we compute the $\Delta \chi^2$ between the best-fit k-essence and $\Lambda$CDM models for each dataset combination considered in this work. These values, along with the corresponding preference in $\sigma$, derived from Wilks' theorem~\cite{Wilks:1938dza}, are reported in Table~\ref{tab:posterior_param_limits}. For the baseline and baseline+Pantheon-Plus dataset we find negligible ($1.5\sigma$ and 1.0$\sigma$, respectively) preference for monodromic k-essence over $\Lambda$CDM. Conversely, for the baseline+DESY5 data, the monodromic k-essence scenario is preferred at 3$\sigma$. This strong dependence on the supernovae dataset underscores the need for consistent supernovae data in order to draw robust conclusions about dynamical dark energy~\cite{Efstathiou:2024xcq}. Finally, as detailed in Appendix~\ref{app:DESI_LRG2}, if we exclude the DESI LRG2 data, this preference drops to $0.3\sigma$, $0.6\sigma$, and $2.4\sigma$, for the baseline, baseline+Pantheon-Plus, and baseline+DESY5 datasets. This indicates that the LRG2 measurements plays a significant role in our monodromic k-essence constraints, particularly in the absence of Type Ia supernovae data.

For comparison, assuming the $w_0$-$w_a$ parametrization, we find $\Delta \chi^2=-8,\,-7\,$ and $-18$ for the baseline, baseline+Pantheon-Plus, and baseline+DESY5 datasets, respectively.\footnote{These $\Delta \chi^2$ values are smaller than those reported in Table VI of the DESI analysis~\cite{DESI:2025zgx} because we use the late-time marginalized CMB likelihood. Importantly, our $\Delta \chi^2$ value for the DESI+$Q_{\rm CMB}$ dataset matches their result using the late-time marginalized CMB likelihood (listed as DESI+($\theta_*\,\omega_{\rm b}\,\omega_{\rm bc})_{\rm CMB}$.} Thus, the physically motivated monodromic dark energy model can achieve $\Delta \chi^2$ improvements comparable to those of the phenomenological $w_0$-$w_a$ parametrization, albeit with a larger number of free parameters. Using Wilks' theorem, these values correspond to a preference of 2.4$\sigma$, 2.3$\sigma$, and 3.8$\sigma$, respectively. This preference drops to 1.4$\sigma$, $1.8\sigma$, and $3.3\sigma$ when the LRG2 data are excluded.

Finally, we turn the parameter constraints into constraints on the redshift evolution of the dark energy equation of state and (generalized) potential $V(z)$, which are shown in Fig.~\ref{fig:reconstruction} (top and bottom panels, respectively), marginalized over all other parameters. We show the results for the baseline (left), baseline+Pantheon-Plus (middle), and baseline+DESY5 (right) datasets. The solid white line shows the posterior mean and the gray lines denote the 68\% and 95\% two-tailed confidence limits. For comparison, we also show the best-fit equation of state in the $w_0$-$w_a$ model for each dataset combination (red dot-dashed). For all datasets, the equation of state roughly oscillates around the cosmlogical constant $w_{\rm DE}=-1$. We note that the constraints here are qualitatively consistent with oscillatory features in the non-parameteric reconstruction from~\citet{DESI:2025wyn} (see also \cite{Zhao:2017cud} for previous eBOSS results). Therefore, the monodromic k-essence scenario is capable of reproducing the oscillatory signatures found in data-driven reconstructions of dark energy.


\section{Conclusions}\label{sec:conclusion}

\noindent In this work, we have presented the first observational constraints on monodromic k-essence --- a physically motivated dynamical dark energy model capable of realizing rapid oscillations in $w_{\rm DE}(z)$ about the phantom divide. Using CMB, BAO, and Type Ia supernovae observations, we constrain the amplitude, frequency, phase, and power-law index of the (generalized) k-essence potential. We find that the combination of CMB and DESI DR2 BAO data are consistent with the standard $\Lambda$CDM model, and adding Pantheon-Plus supernovae measurements strengthens the preference for $\Lambda$CDM over monodromic k-essence. In contrast, including DESY5 supernovae leads to a mild preference ($3\sigma$) for monodromic k-essence. For all dataset combinations considered, the preference for monodromic dark energy (and $w_0$-$w_a$) is partially driven by the DESI LRG2 BAO distance measurements. Finally, we showed that the monodromic k-essence model can fit current BAO and supernovae measurements with $\chi^2$ values comparable to those of the phenomenological $w_0$-$w_a$ model, albeit with two more free parameters (one more in case of fixing $\alpha=0$, as preferred by the data and simplifying the model). 

There are several interesting avenues for future investigation. First, generalizing the monodromic k-essence model studied here to a scenario without gradient instabilities is not only necessary to establish a consistent theory, but would also allow us to include datasets that are sensitive to dark energy perturbations (which are nontrivial due to the small speed of sound in this k-essence scenario), such as the latest high-resolution CMB measurements from the Atacama Cosmology Telescope and South Pole Telescope collaborations~\cite{ACT:2025fju,ACT:2025tim,SPT-3G:2025bzu}, as well as CMB lensing and galaxy clustering data~\cite{ACT:2023kun,DESI:2024jis,SPT-3G:2024atg}.
We present a possible solution to these gradient instabilities in Appendix~\ref{app:kess_stab}, but note that further work is needed. 
See however \cite{2024JCAP...03..032D} for a recent study of galaxy clustering within a dark energy scenario with small speed of sound.

Given the discrepancy between the Pantheon-Plus and DESY5 supernovae constraints on dynamical dark energy~\cite{Efstathiou:2024xcq, DES:2025tir}, and the sensitivity of the constraints on the DESI LRG2 measurements, it will be crucial to reassess the constraints on dynamical dark energy with future BAO and supernovae measurements. Furthermore, future BAO measurements from, \emph{e.g.}, DESI, with finer redshift binning would provide an opportunity to constrain higher frequency oscillations than those explored here. Finally, given the broad parameter space of this model and the fact that the $\Lambda$CDM limit is not uniquely defined, it could be valuable to perform a frequentist analysis, such as a profile likelihood \cite{2025PhRvD.111h3504H,2025arXiv250612004H}, in order to assess the influence of prior volume effects on the constraints presented here.

In conclusion, while this work takes important steps in exploring the parameter space of oscillatory dark energy models, current data do not provide substantial evidence for monodromic dark energy over $\Lambda$CDM. Future supernovae and BAO measurements, together with a deeper theoretical understanding of the stability of dark energy perturbations when crossing the phantom divide in this model, will be essential to assess if oscillations in the dark energy sector have wiggled their way into our Universe.


\acknowledgments
\noindent We thank Laura Herold, Colin Hill, Dragan Huterer, and Eiichiro Komatsu for valuable comments on the manuscript. SG thanks Beatriz Tucci and Fulvio Ferlito for non-monotonous conversations on monodromic dark energy.

SG acknowledges support from the Fulbright U.S. Student Program funding SG's stay at MPA, and NSF grant AST-2307727.
MC acknowledges financial support from Fondazione Angelo Della Riccia during the early stage of the work.
The authors acknowledge the Texas Advanced Computing Center (TACC)\footnote{\href{http://www.tacc.utexas.edu}{http://www.tacc.utexas.edu}} at The University of Texas at Austin for providing computational resources that have contributed to the research results reported within this paper. We acknowledge computing resources from Columbia University's Shared Research Computing Facility project, which is supported by NIH Research Facility Improvement Grant 1G20RR030893-01, and associated funds from the New York State Empire State Development, Division of Science Technology and Innovation (NYSTAR) Contract C090171, both awarded April 15, 2010. 

 \bibliographystyle{apsrev4-1}
\bibliography{biblio.bib}

\phantomsection

\appendix

\begin{figure*}[!t]
\centering
\includegraphics[width=0.99\linewidth]{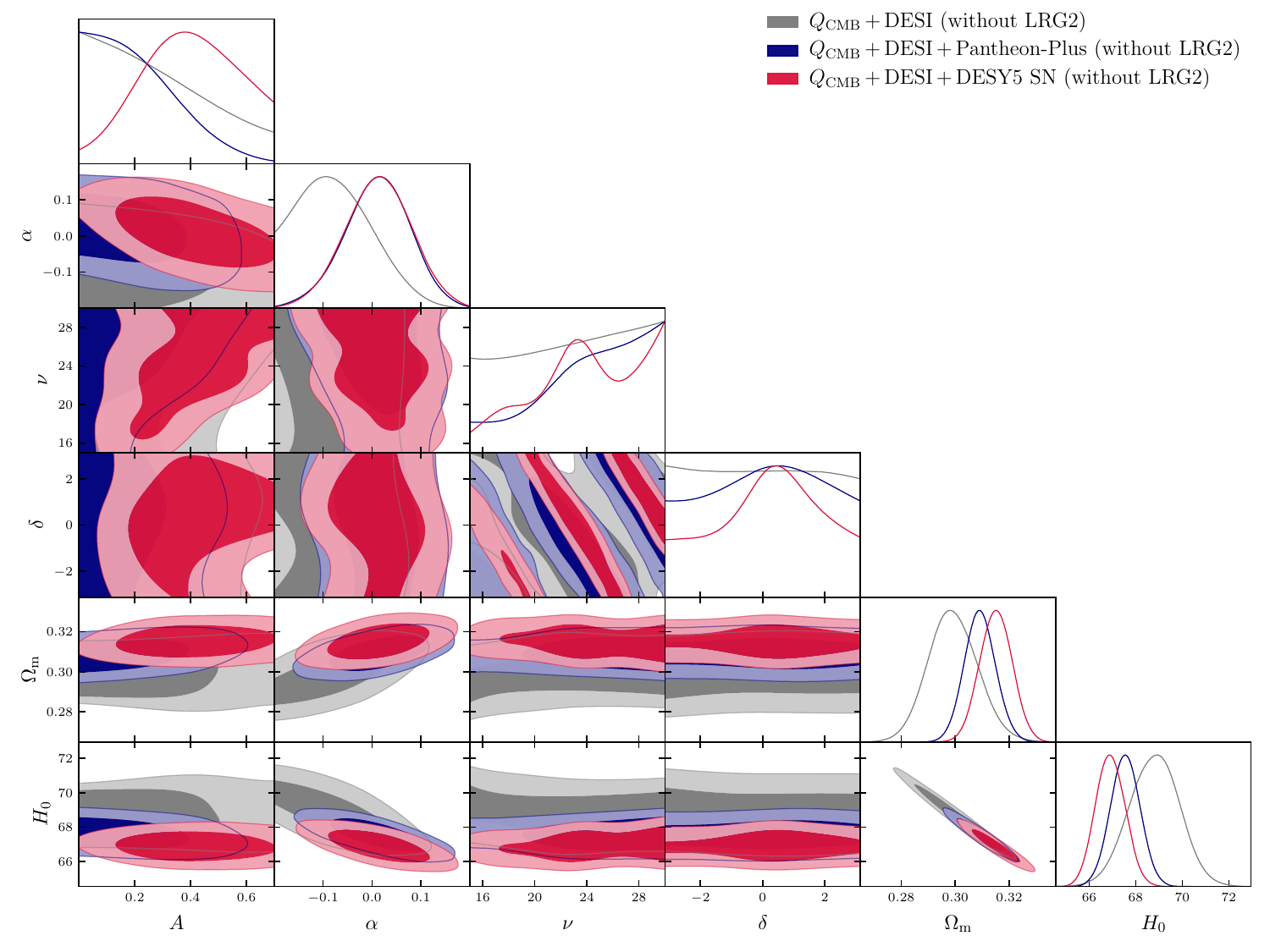}
\caption{Constraints on monodromic k-essence parameters for the dataset combinations considered in this work after excluding the DESI LRG2 BAO distance measurements. Removing the LRG2 measurements significantly broadens the constraints on the frequency (compared to Fig.~\ref{fig:triangle_plot}).}
\label{fig:triangle_no_LRG2}
\end{figure*}
%


\section{On the stability of monodromic k-essence models}\label{app:kess_stab}


\noindent We here recap how, following \citet{Creminelli:2006xe, Creminelli:2008wc} (see also \cite{2024JCAP...03..032D}), adding higher-derivative contributions to the k-essence action in Eqs.~\eqref{eq:MDE_action}--\eqref{eq:MDE_potential} can cure gradient instabilities, and propose a modified k-essence scenario that quantitatively passes this constraint.
In this framework, the monodromic k-essence is an effective field theory that, when crossing the phantom divide line,  should be thought of as a deformation of the ghost condensate \cite{Arkani-Hamed:2003pdi}. Note that this is not the only possibility for stable NEC violations, as other approaches have been proposed based on generalized Galileons \cite{Nicolis:2009qm, Creminelli:2010ba}.

Within the regime of validity of the effective scalar field theory, \emph{i.e.}, on energy scales below the cutoff, the higher-derivative terms yield an upper bound on the growth rate of Jeans and gradient instabilities. Let us denote the energy scale associated with the higher-derivative contributions by $\bar M$, which also defines the cutoff of the effective theory.
In general, $\bar M$ is required to be greater than the minimum scale at which gravity has been probed (cosmologically), that is, $\bar {M} \gtrsim \sqrt{HM_{\rm Pl}}\sim 10^{-3} \rm eV$.

Back-reaction on the dark energy perturbations induced by gravity results in a Jeans instability. Keeping this instability under control, by imposing that the instability rate is smaller than the Hubble rate,\footnote{This constraint is very conservative. As shown in \cite{Arkani-Hamed:2005teg}, much larger instability rates could be experimentally allowed.} implies the following constraint \cite{Creminelli:2008wc}:
\begin{equation}
\bar M \lesssim \left(H M_{\rm Pl}^2\right)^{1/3} \sim 10\:\rm MeV,
\end{equation}
taking $H=H_0$ as the Hubble rate today.

\begin{figure*}[!t]
\centering
\includegraphics[width=0.99\linewidth]{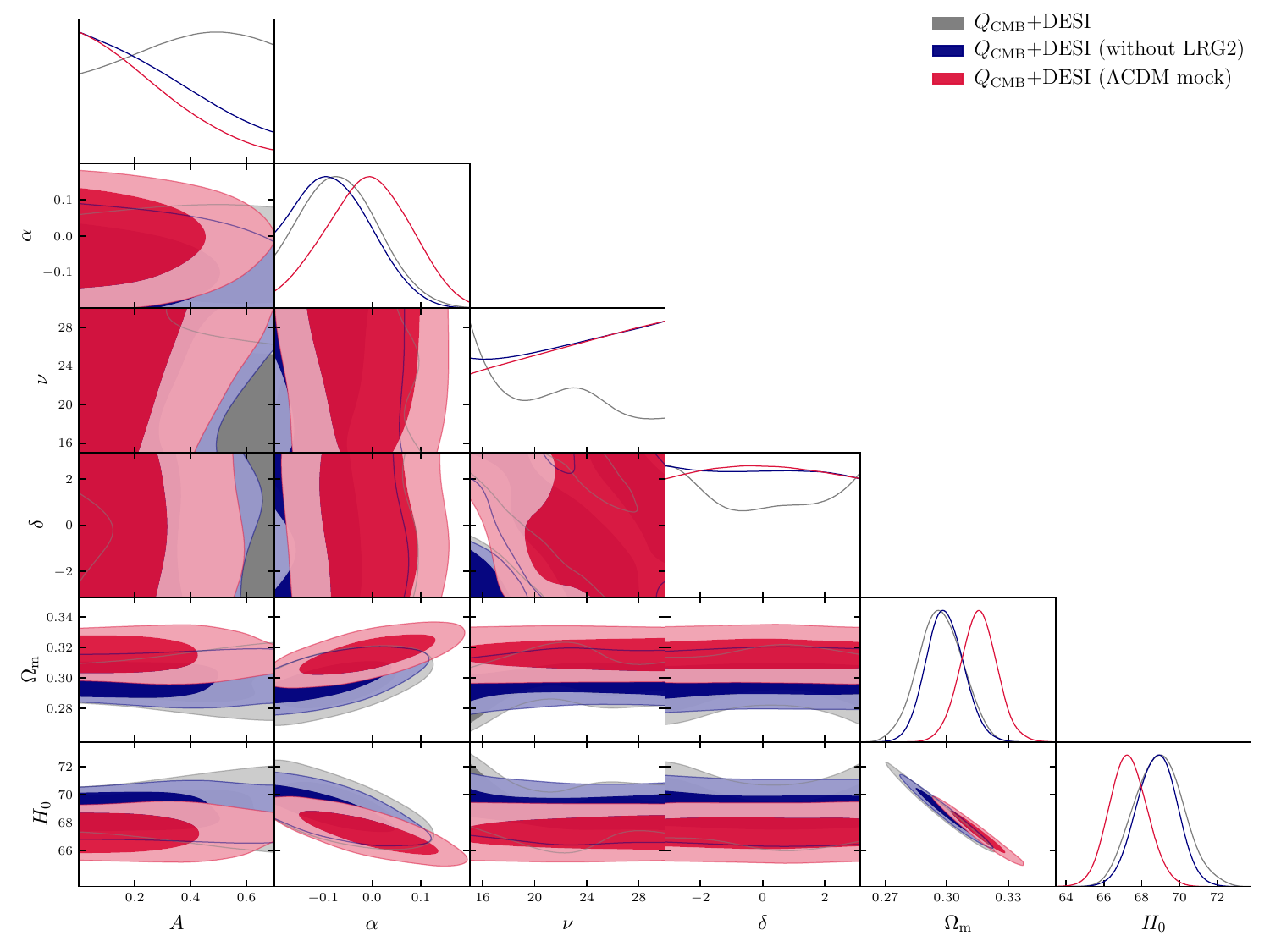}
\caption{Marginalized posterior distributions for the key monodromic dark energy and other cosmological parameters for the baseline dataset with (gray) and without (red) the DESI LRG2 BAO distance measurements. Additionally, to better understand the parameter degeneracies in the monodromic dark energy scenario and to quantify whether the finite redshift binning of the DESI BAO observations leads to a preferred oscillation scale, we show constraints based on a mock $\Lambda$CDM dataset (red) described in the text. }
\label{fig:triangle_plot_mock_data}
\end{figure*}

\renewcommand{\arraystretch}{1.4} 
\begin{table*}[!t]
    \centering
    \scriptsize
     \setlength{\tabcolsep}{2pt} 

        \begin{tabular}{ |c|c|c|c|c|c|c|c| } 
    \hline
    & $A$ & $\alpha$ & $\nu$ & $\delta$ & $\Omega_m$ & $H_0$ [km/s/Mpc] & $\Delta \chi^2$  \\ 
    \hline
    Base: $Q_{\rm CMB}$+DESI 
    & ---$~(0.65)$ 
    & $-0.07^{+0.06}_{-0.10}~(-0.11)$ 
    & ---$~(29.1)$ 
    & ---$~(1.1)$ 
    & $0.299\pm 0.009~(0.294)$ 
    & $68.8\pm 1.1~(69.5)$  
    & -2 \\ 
    \hline
    Base+Pantheon-Plus 
    & $<0.49~(0.26)$ 
    & $0.01^{+0.07}_{-0.06}~(0.01)$ 
    & --- $~(24.8)$ 
    & --- $~(-2.3)$ 
    & $0.309\pm 0.006~(0.312)$ 
    & $67.6\pm 0.6~(67.3)$  
    & -3 \\ 
    \hline
    Base+DESY5 SN 
    & $0.39\pm 0.16~(0.69)$ 
    & $0.01\pm 0.07~(-0.00)$ 
    & ---$~(30.0)$ 
    & ---$~(-0.5)$ 
    & $0.315\pm 0.006~(0.319)$ 
    & $66.9\pm 0.6~(66.6)$  
    & -12  \\ 
    \hline
\end{tabular}

   \caption{Marginalized constraints on monodromic dark energy and other cosmological parameters for the datasets considered in this work after excluding the DESI LRG2 BAO distance measurements. ``---" denotes parameters that are entirely constrained by the prior at $2\sigma.$}
   \label{tab:posterior_param_limits_NO_LRG2}
\end{table*}

\begin{figure*}
  \centering
  \subfloat[Including DESI DR2 LRG2 data.]{%
    \includegraphics[width=0.49\linewidth]{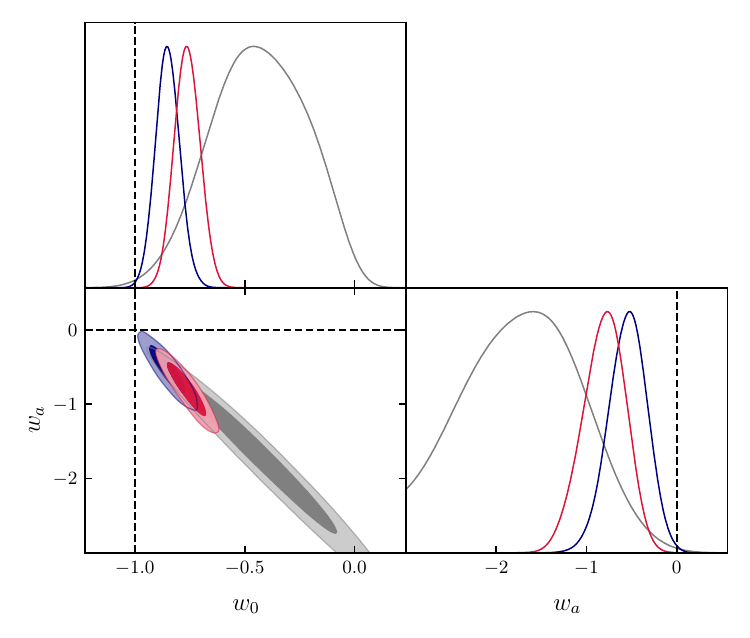}
    \label{fig:sub1}
  }
  \hfill
  \subfloat[Without DESI DR2 LRG2 data.]{%
    \includegraphics[width=0.49\linewidth]{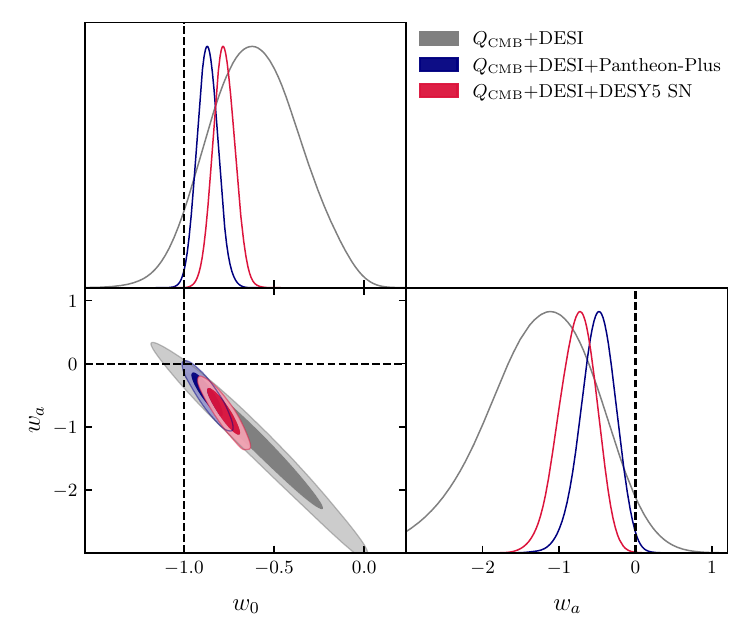}
    \label{fig:sub2}
  }
  \caption{Constraints on $w_0$-$w_a$ models based on the dataset combinations considered in this work. The constraints in the left use the full DESI DR2 BAO dataset, whereas the constraints on the right exclude the parallel and transverse LRG2 BAO distance measurements.}
  \label{fig:LRG2_w0wa}
\end{figure*}

In addition, requiring that the fastest growth rate of dark energy perturbations due to the gradient instabilities is less than the Hubble rate today leads to \cite{Creminelli:2008wc}
\begin{equation}
c_s^2 \approx \frac{\rho_{\rm DE}(1+w_{\rm DE})}{4\bar{M}^4}\gtrsim - \frac{H_0}{\bar M} \sim -10^{-40}\,,
\end{equation}
where we assume the maximum\footnote{Choosing the minimum allowed value for $\bar M$, \emph{i.e.} $\bar M \sim \sqrt{H_0M_{\rm Pl}}$, we get $c_s^2\gtrsim - 10^{-30}$, and  $(1+w_{\rm DE})\Omega_{\rm DE}  \gtrsim    - \frac{4}{3} \sqrt{\frac{H_0}{M_{\rm Pl}}}\sim -10^{-30}$.} allowed value for $\bar M$, that is $\bar M \sim\left(H_0 M_{\rm Pl}^2\right)^{1/3}$. 
 With this choice for $\bar M$, we get the following condition on the equation of state:
\begin{equation}
(1+w_{\rm DE})\Omega_{\rm DE}\gtrsim - \frac{4}{3} \,.
\end{equation}
Thus, the sound speed can drop below zero, but by only a very small quantitative amount, while allowing for interesting phenomenology across the phantom divide line.
The simple k-essence model adopted in this paper has
$c_s^2 \approx (1+w_{\rm DE})/8$, which clearly violates the constraint on $c_s^2$ when $w_{\rm DE}$ drops significantly below $-1$. However, it is not difficult to come up with k-essence-like models that satisfy the constraint. An example is given by
\begin{align}
\label{eq:MDE_stable}
  p(\phi, X) &= \frac{\bar{M}^4}{2}  (2X-1)^2 - F(\phi) + G(\phi) (2X+1) \\
  F(\phi) &= V_0 \left[1 - \tilde{A} \sin(\tilde{\nu} H_0\phi)\right] \\
  G(\phi) & = V_0 \tilde{A} \tilde{\nu} H_0 \cos(\tilde{\nu} H_0\phi).
\end{align}
where $\bar M \sim\left(H_0 M_{\rm Pl}^2\right)^{1/3}$, thus avoiding Jeans instabilities, and $V_0\sim 3 H_0^2 M_{\rm Pl}^2 $ can be adjusted to match the present-day dark energy density similarly to the model considered in the main text.

This model exhibits a sound speed $c_s^2$ that remains above $-10^{-40}$, \emph{i.e.} in the regime where gradient instabilities are stabilized by the higher-derivative contribution, even
when considering parameter values for $\tilde{A}$, $\tilde{\nu}$ that lead to rapid oscillations in $w_{\rm DE}$ with amplitude of order 0.1. In the future, it would be worth performing a background- and perturbation-level analysis of this model as well.

In Eq.~\eqref{eq:MDE_stable}, we have not written the higher-derivative contribution, as its precise form is not important on cosmological scales. Possible choices are \cite{Creminelli:2008wc}:
\begin{align}
\Delta\mathcal{L}_{\rm DE, h.deriv.} =& -\frac{\bar M^2}2 \left[\Box\phi + 3 H(\phi)\right]^2\,  \quad \text{or}
\\
& -\frac{\bar M^3}2 \left[\Box\phi + 3 H(\phi)\right] (2X-1)\,,
\end{align}
where the $H(\phi)$ correction removes the contribution to the background for simplicity.

\begin{figure*}[!t]
\centering
\includegraphics[width=0.99\linewidth]{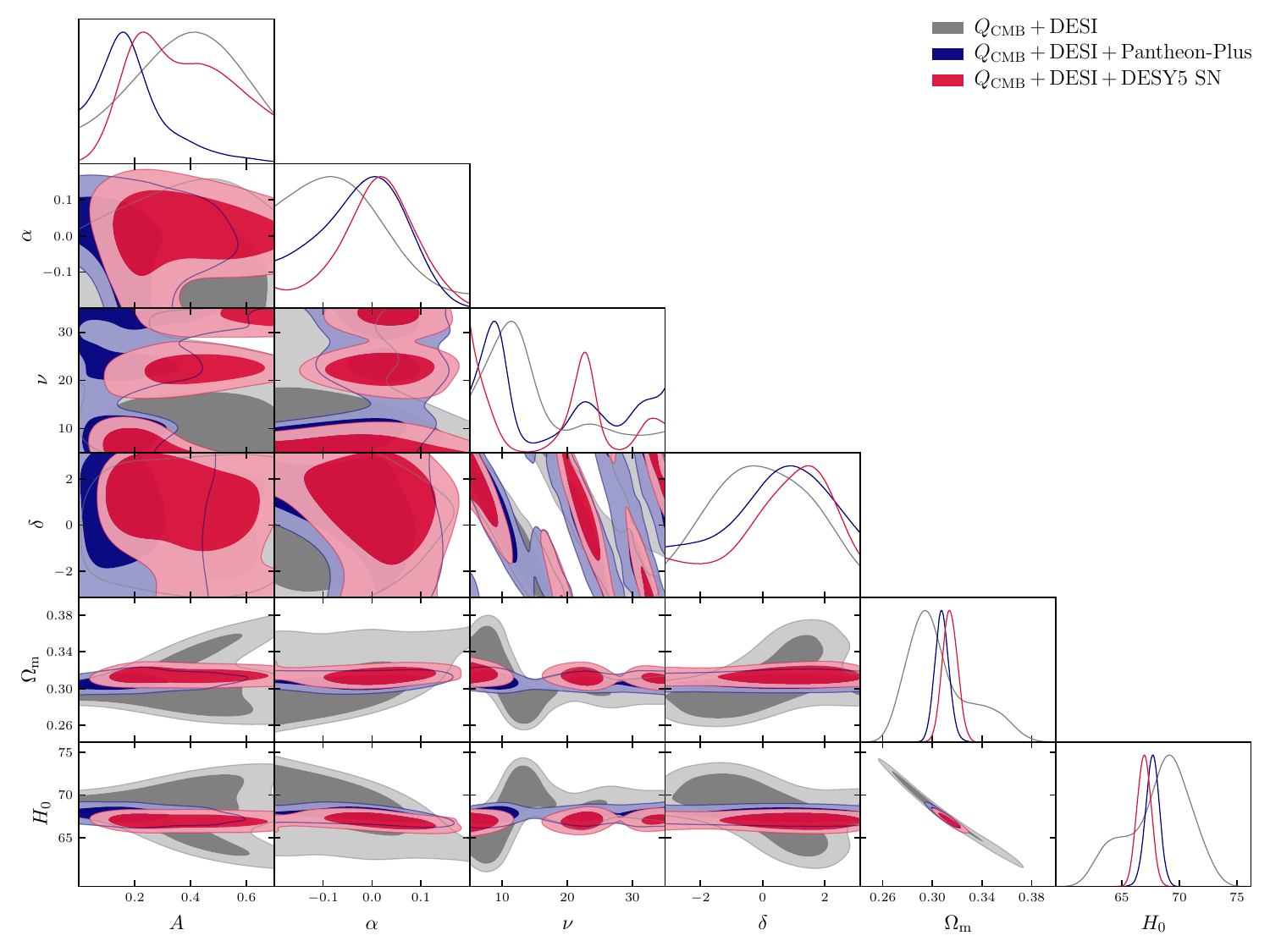}
\caption{Marginalized posterior distributions for the key monodromic dark energy and other cosmological parameters analyzed in this work assuming $5\leq \nu \leq 35$. Constraints are shown for datasets combining \emph{Planck} CMB and DESI BAO (gray), with the addition of Pantheon-Plus supernovae (blue), and DESY5 supernovae (red). }
\label{fig:triangle_plot_extended_nu_prior}
\end{figure*}
\section{Impact of DESI LRG2 Data}\label{app:DESI_LRG2}

\noindent In this appendix, we explore whether the oscillation scales preferred by the DESI DR2 BAO data are influenced by the LRG2 measurements. As detailed in~\citet{DESI:2025zgx}, the DESI LRG2 measurements are in mild tension ($1.5\sigma-2.6\sigma$ depending on assumptions about correlation) with previous measurements from the Sloan Digital Sky Survey (SDSS)~\cite{eBOSS:2020yzd, eBOSS:2020lta,eBOSS:2020hur}. Additionally, the best-fit monodromic dark energy scenarios explored in the main text generally try to fit the dip in the LRG2 parallel BAO distance measurement (see the left panel of Fig.~\ref{fig:BAO_SN_measurement_comparison}). Consequently, we reanalyze the monodromic dark energy scenario without the LRG2 data. 

Fig.~\ref{fig:triangle_no_LRG2} shows the marginalized posteriors for the monodromic dark energy parameters for the datasets considered in this work, excluding the DESI DR2 LRG2 measurements. Compared to our fiducial analysis (Fig.~\ref{fig:triangle_plot}), the constraints on the frequency $\nu$ become significantly weaker.

Fig.~\ref{fig:triangle_plot_mock_data} compares the marginalized posteriors for the $Q_{\rm CMB}+$ DESI BAO dataset combination, with and without the LRG2 data, to those obtained using a mock $\Lambda$CDM BAO dataset. These mock $\Lambda$CDM data consist of parallel, transverse, and isotropic BAO measurements using the same redshift bins and covariance as the combined DESI DR2 BAO dataset, but with central values \emph{fixed} to the predictions for a $\Lambda$CDM cosmology, assuming the central cosmological parameters in the $Q_{\rm CMB}$ multivariate Gaussian likelihood. The constraints on the monodromic dark energy amplitude, frequency, and phase from the mock $\Lambda$CDM dataset (blue) closely resemble those obtained using the DESI observations without the LRG2 data (red).\footnote{The constraint on the power-law index $\alpha$ from the $Q_{\rm CMB}$+DESI dataset (with or without LRG2) is slightly lower than that from the mock data ($\alpha$ = 0), since this parameter can partially compensate for the mild tension in $\Omega_m$ values preferred by the CMB and DESI DR2 BAO under $\Lambda$CDM.}

Finally, we note that this mock data exercise shows that the finite redshift binning of the DESI observables does not lead to a significant preference for a particular oscillation scale, or frequency, if the data are consistent with $\Lambda$CDM.

To quantify the impact of excluding the LRG2 measurements, Table~\ref{tab:posterior_param_limits_NO_LRG2} lists the marginalized posteriors, MAP, and $\Delta \chi^2$ for the analyses without the LRG2 data. The $|\Delta\chi^2|$ are consistently lower than those from the analyses including the LRG2 measurements (Table~\ref{tab:posterior_param_limits}). Indeed, using Wilks' theorem, the preference for monodromic dark energy drops from 1.5$\sigma$, $1.0\sigma$, and $3.0\sigma$, for the $Q_{\rm CMB}$+DESI $Q_{\rm CMB}$+DESI+Pantheon-Plus, and $Q_{\rm CMB}$+DESI+DESY5 datasets, respectively, to $0.3\sigma$, $0.6\sigma$, and $2.4\sigma$.

\begin{figure*}[!t]
\centering
{\includegraphics[width=0.99\linewidth]{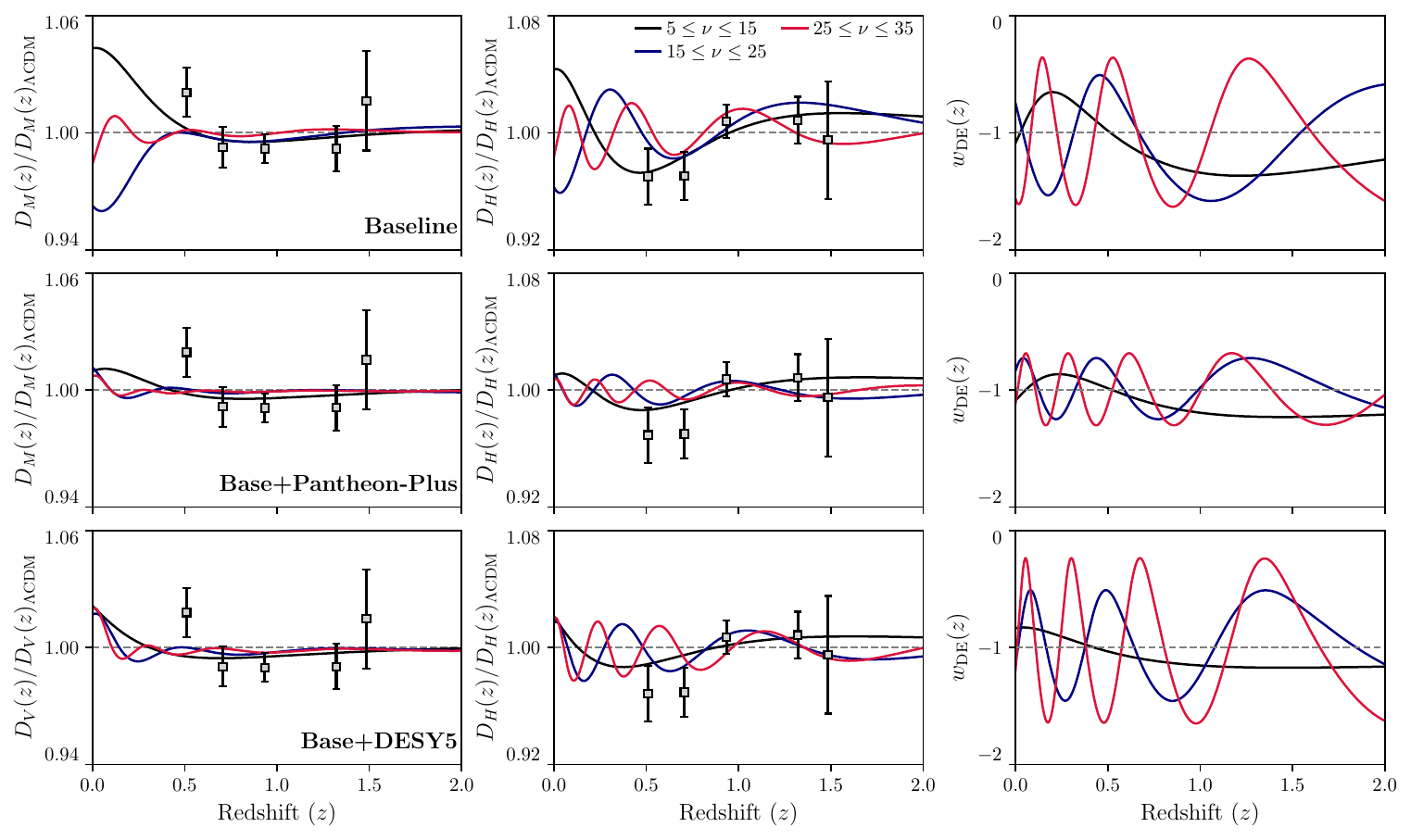}}
\caption{Transverse BAO distance (left), parallel BAO distance (middle), and dark energy equation of state (right) for the best-fit monodromic dark energy model in three different frequency ranges. The top panel shows results based on the baseline dataset. The middle and bottom panels include Pantheon-Plus and DESY5 supernovae, respectively. The BAO distances are shown relative to the best-fit $\Lambda$CDM cosmology for the particular dataset combination. Squares denote DESI DR2 measurements.}
\label{fig:BAO_SN_measurement_comparison_wide_prior}
\end{figure*}

For comparison, we repeat the exercise of removing the LRG2 data for the $w_0$-$w_a$ model. Fig.~\ref{fig:LRG2_w0wa} shows the marginalized posteriors on $w_0$ and $w_a$ with (left) and without (right) the LRG2 data. Removing the LRG2 data shifts all contours towards the $\Lambda$CDM limit ($w_0=-1$ and $w_a=0$). The $\Delta \chi^2$ values for $Q_{\rm CMB}+$DESI, $Q_{\rm CMB}+$DESI+Pantheon-Plus, and $Q_{\rm CMB}+$DESI+DESY5 SN are -4, -5, -14, corresponding to a $1.4\sigma$, $1.8\sigma$, and $3.3\sigma$ preference for this model, respectively. These are lower than the $2.4\sigma$, $2.3\sigma$, and $3.8\sigma$ preference for $w_0$-$w_a$ based on the analyses including LRG2. 

In conclusion, our results show that the mild preference for dynamical dark energy when combining CMB and DESI BAO data is partially driven by the LRG2 bin.\footnote{As mentioned in the main text, the statistical preferences quoted in this work are somewhat lower than the main results of the DESI analyses~\cite{DESI:2025zgx}, which use a full CMB likelihood (including lensing).} We emphasize that removing an individual redshift bin, as we have done here, is somewhat extreme, especially given that the low redshift LRG2 data are particularly important for constraining dynamical dark energy. Future DESI BAO measurements in this redshift range will be critical for assessing the viability of dynamical dark energy, especially in analyses that do not include supernovae data.

\section{Constraints assuming a broader prior on frequency}\label{app:relaxed_frequency_prior}

\noindent In the main text, we imposed an informative prior on the monodromic dark energy frequency ($15 \leq \nu \leq 30$) to restrict our analysis to a regime with several oscillations for $z\lesssim 2$, while avoiding frequencies that are too high to resolve given the redshift binning of the DESI DR2 data. Here, we relax this assumption and present constraints using a broader prior, $5 \leq \nu \leq 35$. In this section, we adopt a relaxed convergence criterion with $R-1<0.05.$

Fig.~\ref{fig:triangle_plot_extended_nu_prior} shows the marginalized posteriors for the runs with a broad frequency prior. Relaxing the frequency prior leads to a multimodal posterior, indicating that multiple oscillation frequencies provide comparable fits to the data.

To further investigate this multimodality, we compute the best-fit (MAP) cosmological parameters within three frequency intervals: $5\leq \nu \leq 15$, $15 \leq \nu \leq 25$, and $25 \leq \nu \leq 35$, for each dataset combination. Fig.~\ref{fig:BAO_SN_measurement_comparison_wide_prior} compares the transverse BAO distance (left), parallel BAO distance (middle), and dark energy equation of state (right) for the best-fit monodromic dark energy models in each frequency interval. The top panel shows the results form the \emph{baseline} dataset. The overall best-fit model in the extended prior range has $\nu \approx 9$, but only improves the fit by $\Delta \chi^2 \approx -1$ compared to the model with $\nu \approx 15$ in Table~\ref{tab:posterior_param_limits}. In contrast, models with $\nu>25$ are generally disfavored by the baseline dataset, with the best-fit cosmology in that range yielding a $\Delta \chi^2\approx4$ worse fit than the best-fit over the full extended frequency range. Note that, in all cases, the best-fit parallel BAO distances for the baseline dataset have local minima near the LRG2 measurement.

The middle and bottom panels of Fig.~\ref{fig:BAO_SN_measurement_comparison_wide_prior} show the best-fit predictions for the combination of the baseline dataset with Pantheon-Plus and DESY5 Type Ia supernovae, respectively. For the Pantheon-Plus dataset combination, the best-fit cosmology with $\nu \approx 8$ improves upon the $\nu \approx 22$ best-fit in Table~\ref{tab:posterior_param_limits} by $\Delta \chi^2\approx -4$. Conversely, the best-fit cosmology with $\nu>25$ provides a worse fit to the data than the $\nu \approx 22$ best-fit by $\Delta \chi^2\approx 1.$ For the DESY5 dataset combination, the overall best-fit with $\nu \approx 5$ improves upon the $\nu \approx 22$ best-fit model studied in the main text by $\Delta \chi^2 \approx -1$. The best-fit model in the high frequency bin ($\nu \approx32$) is only $\Delta \chi^2\approx 2$ worse than the overall best fit.

In summary, current BAO and supernovae datasets are consistent with a broad range of frequencies in the monodromic dark energy scenario, including both lower and higher values than those explored in the main text. Additional data will be necessary to better constrain this extended parameter space.

\end{document}